\documentclass[fleqn]{article}
\usepackage{longtable}
\usepackage{graphicx}

\begin {document}
\begin{flushleft}
{\LARGE
{\bf Comment on ``Atomic structure and excitation cross section by electron impact of tungsten ions, W$^{38+}$" by El-Maaref {\it et al}\,   [J. Phys. B  52(2019) 065202]}
}\\

\vspace{1.5 cm}

{\bf {Kanti  M  ~Aggarwal}}\\ 

\vspace*{1.0cm}

Astrophysics Research Centre, School of Mathematics and Physics, Queen's University Belfast, \\Belfast BT7 1NN, Northern Ireland, UK\\ 
\vspace*{0.5 cm} 

e-mail: K.Aggarwal@qub.ac.uk \\

\vspace*{0.20cm}

Received: 9 September 2019

\vspace*{1.0 cm}

\vspace*{1.0 cm}
\vspace*{1.0 cm}

\hrule

\vspace{0.5 cm}

\end{flushleft}

\clearpage


\begin{abstract}

Recently, El-Maaref {\it et al}\, [J. Phys. B  52(2019) 065202] have reported results for  energy levels, radiative rates and collision strengths  ($\Omega$)  for  some transitions  of Kr-like W~XXXIX. For the calculations of these parameters they have adopted several codes, including GRASP, DARC and FAC. For energy levels they have shown discrepancies of up to 1.64~Ryd between the GRASP and FAC calculations, whereas for $\Omega$ the differences between the DARC and FAC results are over an order of magnitude for a few transitions. In addition, for several transitions there is an anomaly in the behaviour of $\Omega$ between the two sets of calculations. In this comment, we demonstrate  that their results from both codes are incorrect, and hence cannot be relied upon.

\end{abstract}

\clearpage

\section{Introduction}

In a recent paper, El-Maaref {\it et al}\, (2019) have reported results for energy levels, radiative rates and collision strengths  ($\Omega$)  for a few transitions  of Kr-like W~XXXIX, which is an important ion particularly for the studies of fusion plasmas. For the calculations of atomic structure, i.e. for the determination of energy levels, radiative rates (A-values) and  oscillator strengths (f-values), they have adopted several codes, namely the General-purpose Relativistic Atomic Structure Package (GRASP: {\tt http://amdpp.phys.strath.ac.uk/UK\_APAP/codes.html}), the Flexible Atomic Code (FAC: {\tt https://www-amdis.iaea.org/FAC/}) and AutoStructure (AS: Badnell 1997). This was for making inter-comparisons among several sets of calculations and to make an assessment of accuracy. However, the discrepancies shown in energies for a few levels are very large, i.e. up to 1.64~Ryd -- see for example level 127 in their table~1. In our long experience of working with these codes and for a wide range of ions, including those of tungsten (Aggarwal and Keenan 2016), such large discrepancies have not been observed, and therefore their results appear to be suspicious. Similarly, for the calculations of $\Omega$ they have adopted the $R$-matrix (the Dirac Atomic $R$-matrix Code, DARC: {\tt http://amdpp.phys.strath.ac.uk/UK\_APAP/codes.html}) and the distorted-wave (DW) methods (FAC). Both of these codes are fully relativistic and the main difference between the two, for a highly ionised system such as W~XXXIX, is in the calculations of closed-channel (Feshbach) resonances in the former but not in the latter. Again, our past experience on many ions, including the one of tungsten (W~LXVI: Aggarwal 2016), shows that generally the background values of collision strengths ($\Omega_B$) are comparable between the two calculations for a majority of transitions, and over a large range of energies, provided similar number of configurations and their configuration state functions (CSF), or correspondingly the number of levels, are included in both. However, El-Maaref {\it et al}\, (2019) have noted large discrepancies in $\Omega$, in both the magnitude as well as the behaviour, between the two calculations -- see for example, transitions 1--2 and 1--5 in their figure~1. Therefore, for the benefit of the readers, we demonstrate in this short comment that their calculations are incorrect from all codes, and if the calculations are performed correctly  with care then the discrepancies are not as striking as shown by them. 

\section {Energy levels and radiative rates}

For calculating energy levels, El-Maaref {\it et al}\, (2019) have chosen two models consisting of: (i) 25 levels of the (4s$^2$)4p$^6$, 4p$^5$4d and 4p$^5$4f, and (ii) 357 levels, the additional ones arising from the inclusion of the 4p$^4$4d$^2$, 4p$^4$4f$^2$ and 4p$^5$5$\ell$ configurations. As already stated, they have performed three calculations with these models with the GRASP, FAC and AS codes, and the discrepancies shown for some of the levels are up to 1.64~Ryd -- see in particular level 127 in their table~1. We have performed two calculations with GRASP and FAC for both models, but in table~1 compare the energies for only the larger model, and for the same levels as listed in their table~1. The orderings of levels are the same in both calculations and (nearly) match with those of El-Maaref {\it et al}.\,  However, there are two major differences between our and their calculations. Firstly, they have incorrectly identified the configurations for some of the levels, such as 34, 52, 61, 124 and 127, which belong to the 4p$^4$4d$^2$ configuration in stead of 4p$^5$4f, as listed by them. In fact, this is the main reason that the discrepancies shown by them between the smaller and larger model calculations are are up to 1.64~Ryd, as the correspondence of the levels between the two is incorrect. Secondly, the differences between our calculations with GRASP and FAC (GRASP2 and FAC2) are `expectedly' insignificant, whereas in their work (GRASP1 and FAC1) these are up to 0.7~Ryd for most levels, and energies with GRASP are invariably higher, in spite of using the same {\em configuration interaction}  (CI) in both codes. Finally, the differences between the GRASP1 and GRASP2 energies are generally within $\sim$0.15~Ryd, but between FAC1 and FAC2 are up to $\sim$0.5~Ryd, and their values are invariably {\em lower}.

\begin{table}
\caption{Comparison of energy levels (in Ryd) for Kr-like W~XXXIX.} 
\begin{tabular}{rllrrrrrr} \hline
\\
Index &   Configuration                    & Level      &  GRASP1    & FAC1   &    GRASP2  & FAC2  \\ 
\hline \\												       
1   &	4p$^6$  			   & $^1$S$_0$    &    0.000   &   0.000  &  0.0000  &  0.0000 \\
2   &	4p$^5$4d			   & $^3$P$^o_0$  &    11.443  &  10.904  & 11.4184  & 11.4110 \\
3   &	4p$^5$4d			   & $^3$P$^o_1$  &    11.789  &  11.242  & 11.7577  & 11.7489 \\
4   &	4p$^5$4d			   & $^3$F$^o_3$  &    12.033  &  11.480  & 11.9984  & 11.9863 \\
5   &	4p$^5$4d			   & $^3$D$^o_2$  &    12.123  &  11.574  & 12.0935  & 12.0814 \\
6   &	4p$^5$4d			   & $^3$F$^o_4$  &    13.215  &  12.638  & 13.1466  & 13.1414 \\
7   &	4p$^5$4d			   & $^1$D$^o_2$  &    13.376  &  12.800  & 13.3137  & 13.3074 \\
8   &	4p$^5$4d			   & $^3$D$^o_3$  &    13.789  &  13.209  & 13.7273  & 13.7176 \\
9   &	4p$^5$4d			   & $^3$D$^o_1$  &    14.967  &  14.382  & 14.9056  & 14.8898 \\
10  &	4p$^5$4d			   & $^3$F$^o_2$  &    18.713  &  18.107  & 18.6037  & 18.6114 \\
11  &	4p$^5$4d			   & $^3$P$^o_2$  &    20.190  &  19.562  & 20.0555  & 20.0676 \\
12  &	4p$^5$4d			   & $^1$P$^o_1$  &    20.380  &  19.760  & 20.2773  & 20.2735 \\
13  &	4p$^5$4d			   & $^1$F$^o_3$  &    20.394  &  19.763  & 20.2544  & 20.2654 \\
14  &	4p$^4$($^3$P$_2$)4d$^2$($^3$F$_2$) & $^5$D$_0$    &    22.703  &  22.096  & 22.6467  & 22.6245 \\
15  &	4p$^4$($^3$P$_2$)4d$^2$($^3$F$_2$) & $^5$D$_1$    &    22.970  &  22.360  & 22.9081  & 22.8850 \\
16  &	4p$^4$($^3$P$_2$)4d$^2$($^3$F$_2$) & $^5$F$_2$    &    23.093  &  22.477  & 23.0256  & 23.0021 \\
17  &	4p$^4$($^3$P$_2$)4d$^2$($^3$F$_2$) & $^5$F$_3$    &    23.122  &  22.507  & 23.0568  & 23.0313 \\
18  &	4p$^4$($^3$P$_2$)4d$^2$($^3$F$_2$) & $^5$G$_4$    &    23.333  &  22.710  & 23.2627  & 23.2352 \\
19  &	4p$^4$($^3$P$_2$)4d$^2$($^3$F$_2$) & $^3$F$_2$    &    23.558  &  22.941  & 23.4934  & 23.4655 \\
20  &	4p$^4$($^1$S$_0$(4d$^2$($^3$F$_2$) & $^3$F$_2$    &    23.998  &  23.377  & 23.9335  & 23.9021 \\
21  &	4p$^4$($^3$P$_2$)4d$^2$($^3$F$_2$) & $^5$F$_4$    &    24.302  &  23.662  & 24.2059  & 24.1874 \\
22  &	4p$^4$($^3$P$_2$)4d$^2$($^3$P$_2$) & $^5$S$_2$    &    24.437  &  23.790  & 24.3447  & 24.3259 \\
23  &	4p$^4$($^3$P$_2$)4d$^2$($^3$F$_2$) & $^5$G$_5$    &    24.438  &  23.801  & 24.3337  & 24.3150 \\
24  &	4p$^4$($^3$P$_2$)4d$^2$($^3$F$_2$) & $^5$D$_3$    &    24.480  &  23.840  & 24.3841  & 24.3651 \\
25  &	4p$^4$($^1$S$_0$)4d$^2$($^3$P$_2$) & $^3$P$_0$    &    24.520  &  23.892  & 24.4587  & 24.4235 \\
26  &	4p$^4$($^3$P$_2$)4d$^2$($^3$P$_2$) & $^5$P$_3$    &    24.645  &  24.005  & 24.5508  & 24.5304 \\
27  &	4p$^4$($^3$P$_2$)4d$^2$($^3$P$_2$) & $^5$D$_4$    &    24.652  &  24.005  & 24.5500  & 24.5299 \\
28  &	4p$^4$($^3$P$_2$)4d$^2$($^1$G$_2$) & $^3$G$_5$    &    24.759  &  24.106  & 24.6521  & 24.6305 \\
29  &	4p$^4$($^3$P$_2$)4d$^2$($^3$F$_2$) & $^3$D$_1$    &    24.855  &  24.216  & 24.7607  & 24.7408 \\
30  &	4p$^4$($^3$P$_2$)4d$^2$($^3$F$_2$) & $^5$G$_6$    &    24.910  &  24.251  & 24.7983  & 24.7759 \\
31  &	4p$^4$($^3$P$_2$)4d$^2$($^1$G$_2$) & $^3$F$_2$    &    25.060  &  24.416  & 24.9647  & 24.9406 \\
32  &	4p$^4$($^1$S$_0$)4d$^2$($^3$F$_2$) & $^3$F$_3$    &    25.371  &  24.725  & 25.2765  & 25.2498 \\
33  &	4p$^4$($^3$P$_2$)4d$^2$($^3$F$_2$) & $^3$D$_2$    &    25.597  &  24.951  & 25.4964  & 25.4750 \\
34  &	4p$^4$($^3$P$_2$)4d$^2$($^1$G$_2$) & $^3$G$_4$    &    25.658  &  25.020  & 25.5575  & 25.5415 \\
35  &	4p$^4$($^1$D$_2$)4d$^2$($^3$P$_2$) & $^3$D$_1$    &    25.689  &  25.043  & 25.5963  & 25.5680 \\
49  &	4p$^5$4f			   & $^3$D$_1$    &    27.067  &  26.455  & 26.9730  & 26.9689 \\
52  &	4p$^4$($^3$P$_2$)4d$^2$($^1$S$_0$) & $^3$P$_2$    &    27.437  &  26.789  & 27.3180  & 27.3039 \\
56  &	4p$^5$4f			   & $^3$D$_3$    &    27.770  &  27.133  & 27.6547  & 27.6483 \\
57  &	4p$^5$4f			   & $^3$G$_5$    &    27.892  &  27.241  & 27.7671  & 27.7570 \\
61  &	4p$^4$($^1$S$_0$)4d$^2$($^1$S$_0$) & $^1$S$_0$    &    28.044  &  27.387  & 27.9279  & 27.8935 \\
62  &	4p$^5$4f			   & $^1$F$_3$    &    28.166  &  27.515  & 28.0512  & 28.0331 \\
63  &	4p$^5$4f			   & $^3$F$_4$    &    28.377  &  27.727  & 28.2575  & 28.2437 \\
\\ \hline
\end{tabular} 
\end{table}

\setcounter{table}{0}
\begin{table}
\caption{continued.} 
\begin{tabular}{rllrrrrrr} \hline
\\
Index &   Configuration                    & Level      &  GRASP1    & FAC1   &    GRASP2  & FAC2  \\ 
\hline \\

112 &	4p$^5$4f			   & $^3$F$_2$  &    33.688  &  32.990  & 33.5191  & 33.5116 \\
122 &	4p$^5$4f			   & $^3$G$_3$  &    35.120  &  34.424  & 34.9433  & 34.9384 \\
124 &	4p$^4$($^3$P$_2$)4d$^2$($^3$P$_2$) & $^1$S$_0$  &    35.340  &  34.526  & 35.1746  & 35.1403 \\
127 &	4p$^4$($^1$D$_2$)4d$^2$($^1$G$_2$) & $^1$D$_2$  &    35.873  &  35.161  & 35.7008  & 35.6786 \\
\\ \hline
\end{tabular} 

\begin{flushleft}
{\small
GRASP1: earlier calculations of El-Maaref {\it et al}\ (2019) with  the {\sc grasp} code for 357 levels \\
GRASP2: present calculations with  the {\sc grasp} code for 357 levels \\
FAC1: earlier calculations of El-Maaref {\it et al}\ (2019) with  the {\sc fac} code for 357 levels \\
FAC2: present calculations with  the {\sc fac} code for 357 levels \\
}
\end{flushleft}
\end{table}

In conclusion, we will like to state that there is scope for improvement in the calculated energy levels of El-Maaref {\it et al}\, (2019), but their identification of the configuration is incorrect for several levels, and this is the main reason for the large discrepancies shown by them for different models and codes. Finally, El-Maaref {\it et al}\,  have only provided the J values for the levels but we have also listed the LSJ$^{\pi}$ designations in table~1, which may be helpful for further comparisons. Without this there is no distinction among levels of a J value belonging to the same configuration. For example, there are six J = 2 levels for the 4p$^4$4d$^2$ configuration shown in their table~1, i.e. levels 16, 19, 20, 22, 31, and 33. However, these level designations should be used with caution as these may not always be definitive (and are only for guidance), because some of them are highly mixed. As an example, level 127 is a mixture of 0.33~4p$^5$4f~$^3$F$_2$ (22), 0.43~4p$^5$4f~$^1$D$_2$~(24), 0.21~4p$^4$4d$^2$($^3$F$_2$)~$^3$F$_2$~(82), 0.34~4p$^4$4d$^2$($^3$F$_2$)~$^1$D$_2$~(84), 0.23~4p$^4$4d$^2$($^1$G$_2$)~$^3$F$_2$~(91), and 0.48~4p$^4$4d$^2$($^1$G$_2$)~$^1$D$_2$~(102). Hence, this level (and many more) is a mixture of several J levels from two different configurations, which makes it difficult to provide a unique identification for all levels, and this is a general atomic physics problem in all codes and methods. However, the J$^{\pi}$ values and their orderings are definitive and can be used with confidence.

For the A-values, there are no (major) discrepancies between our calculations and those of El-Maaref {\it et al}\, (2019),\, although differences for the 3--25 transition (f = 1.3$\times$10$^{-4}$) are significant (70\%), because their value with GRASP is 2.94$\times$10$^8$ whereas ours is 5.00$\times$10$^8$~s$^{-1}$, matching well with the corresponding results with FAC, their being 5.24$\times$10$^8$ and ours 4.19$\times$10$^8$~s$^{-1}$. This difference  can also not be attributed to a typing mistake because all three results (from GRASP, FAC and AS) listed in their table~2 differ by up to a factor of four. Since this is weak transition such differences may appear sometimes among different calculations.

\begin{figure*}
\includegraphics[angle=-90,width=0.9\textwidth]{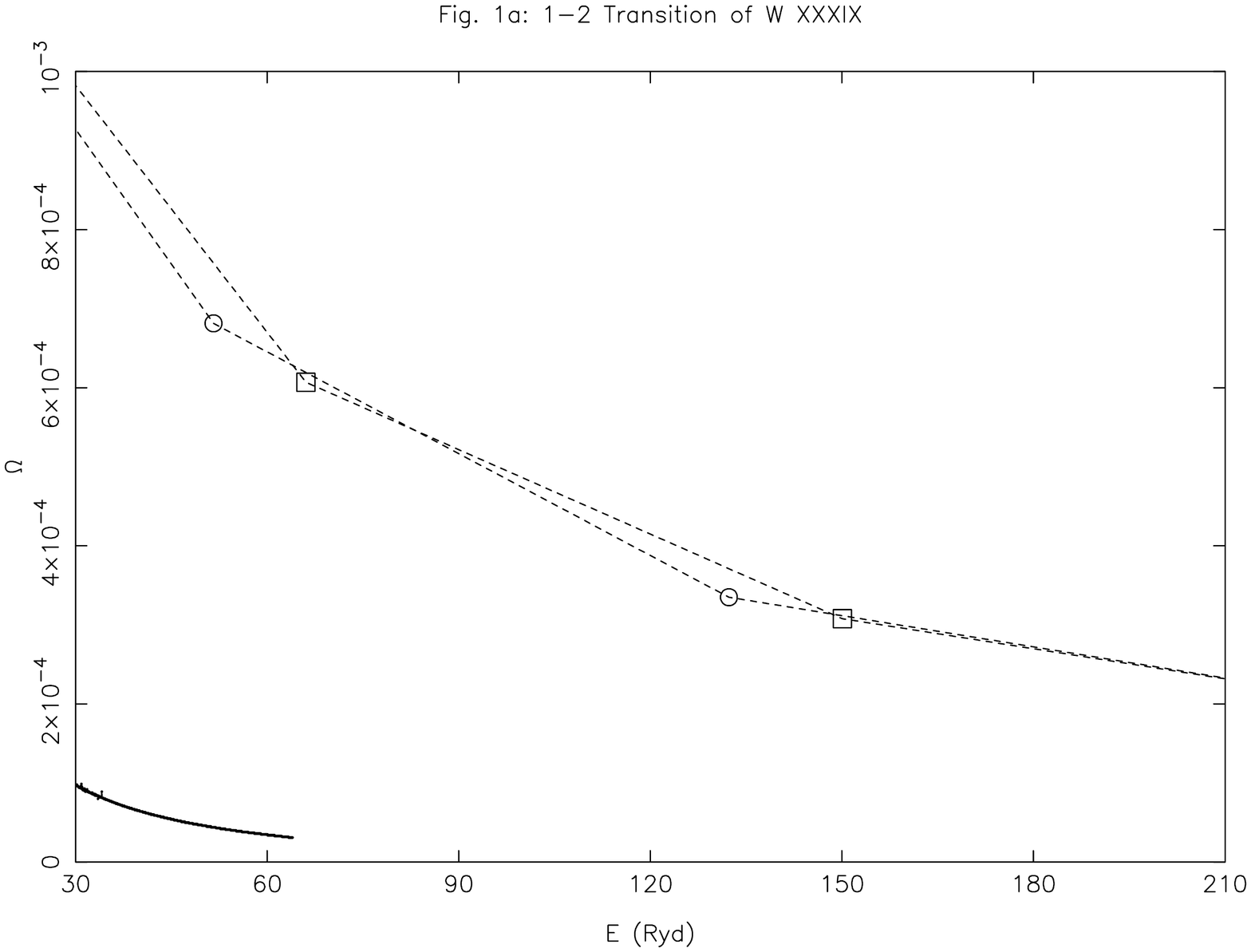}
 \vspace{-1.5cm}
 \end{figure*}
 
\setcounter{figure}{0}
 \begin{figure*}
\includegraphics[angle=-90,width=0.9\textwidth]{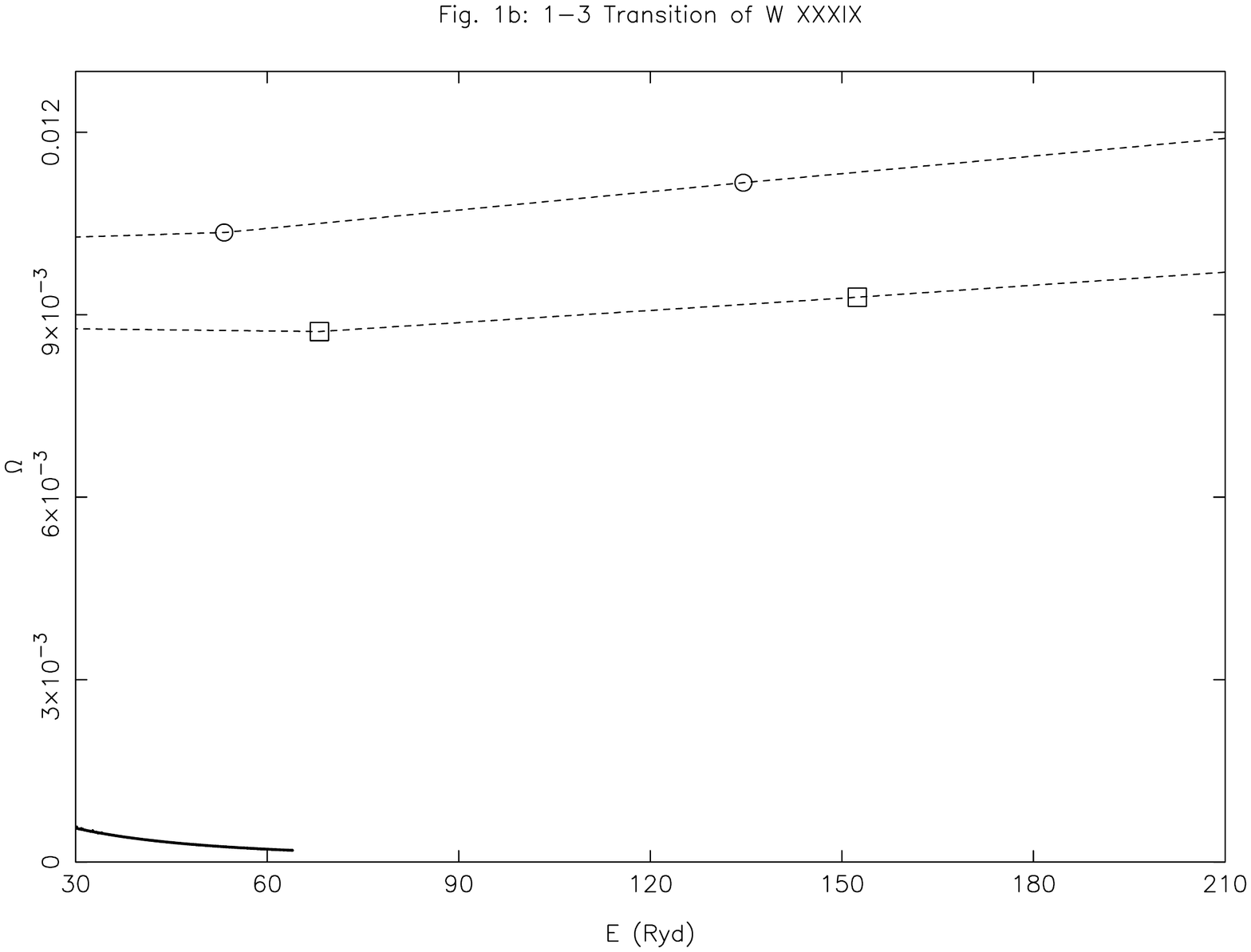}
 \vspace{-1.5cm}
 \end{figure*}
 
 \setcounter{figure}{0}
\begin{figure*}
\includegraphics[angle=-90,width=0.9\textwidth]{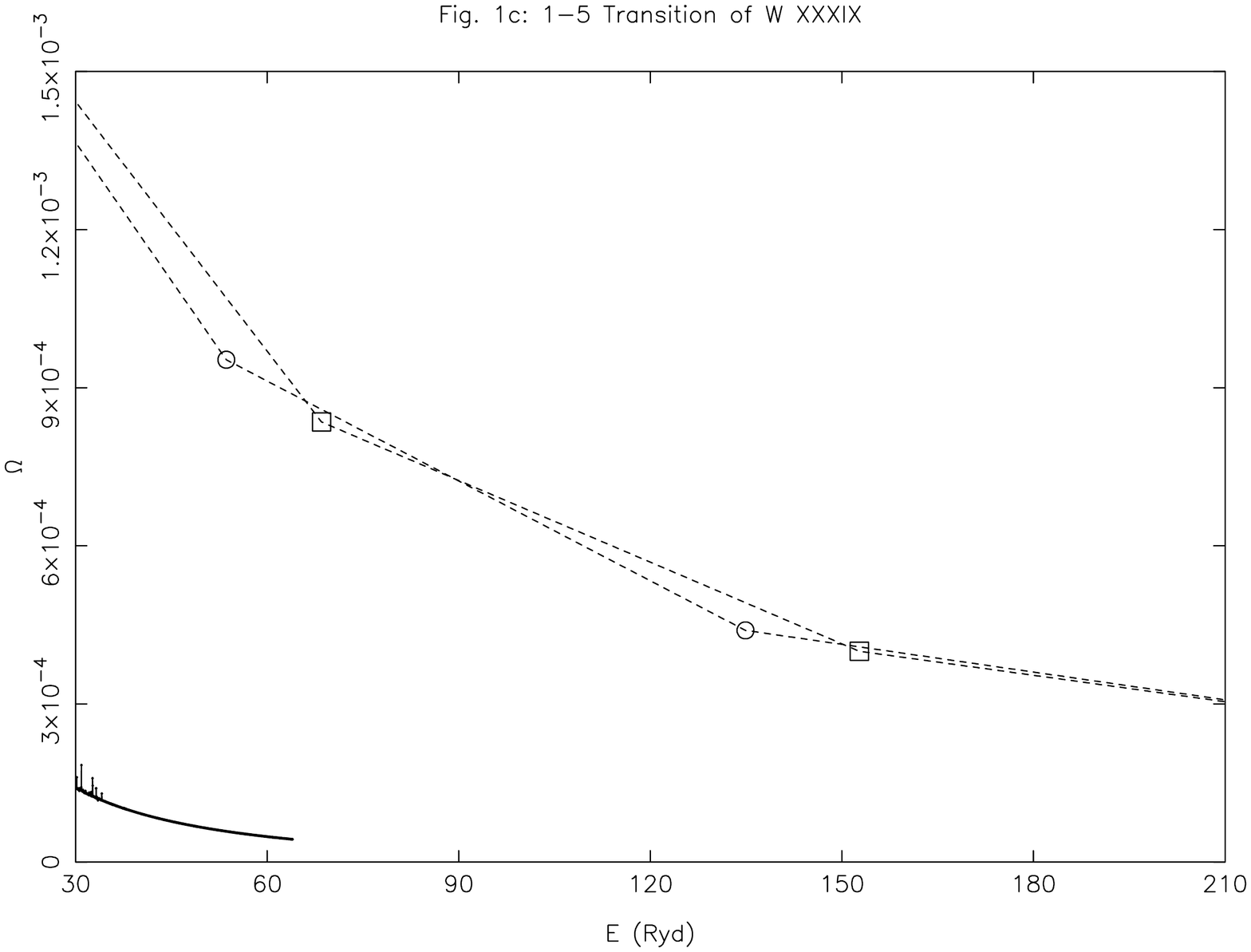}
 \vspace{-1.5cm}
\caption{Comparison of DARC and FAC values of $\Omega$ for  the (a)  1--2 (4p$^6$~$^1$S$_0$ -- 4p$^5$4d~$^3$P$^o_0$), (b)  1--3 (4p$^6$~$^1$S$_0$ -- 4p$^5$4d~$^3$P$^o_1$) and (c) 1--5 (4p$^6$~$^1$S$_0$ -- 4p$^5$4d~$^3$D$^o_2$) transitions of W~XXXIX. Continuous curves: earlier results of El-Maaref {\it et al}\, (2019) with DARC, broken curves: present results  with FAC, circles: 25 levels model and squares: 357 levels model.}
 \end{figure*} 
 
 \section {Collision strengths}

El-Maaref {\it et al}\, (2019) have performed two sets of calculations with DARC for $\Omega$, adopting the smaller and larger models with 25 and 307 (out of 357) levels, listed in section~2. They have resolved resonances in the thresholds region with an energy mesh of 0.02~Ryd, and have calculated results up to an energy of 64~Ryd, which does not even cover the entire thresholds region as it extends to $\sim$80~Ryd.  They have also performed similar calculations, although without resonances, with the DW method as implemented in the AS and FAC codes, and have shown comparisons for several transitions in their figure~1. However, there are three obvious errors in those comparisons. Firstly, for {\em inelastic} transitions $\Omega$ results are not possible for energies {\em below} thresholds, as shown by them. This has happened because DARC lists results  w.r.t. {\em incident} energies, whereas the DW codes list w.r.t. excited ones, and they have not realised this difference. Secondly, for two transitions, namely 1--2 (4p$^6$~$^1$S$_0$ -- 4p$^5$4d~$^3$P$^o_0$) and 1--5 (4p$^6$~$^1$S$_0$ -- 4p$^5$4d~$^3$D$^o_2$), they have shown a sudden {\em jump} in $\Omega$s at energies above $\sim$25~Ryd, and have therefore concluded that the DW results may be inaccurate. However, this kind of jump neither happens in the calculations nor can be explained. Thirdly, they are unable to distinguish between collision strengths ($\Omega$) and cross-sections ($\sigma$), in spite of giving relationship between the two in their eq. (8). This is because in their figures~2, 3 and 4 (and in the related text) they are plotting $\Omega$ (a dimensionless quantity)  but describing $\sigma$. Apart from these obvious errors there are others which are more important, as discussed below.

\begin{figure*}
\includegraphics[angle=-90,width=0.9\textwidth]{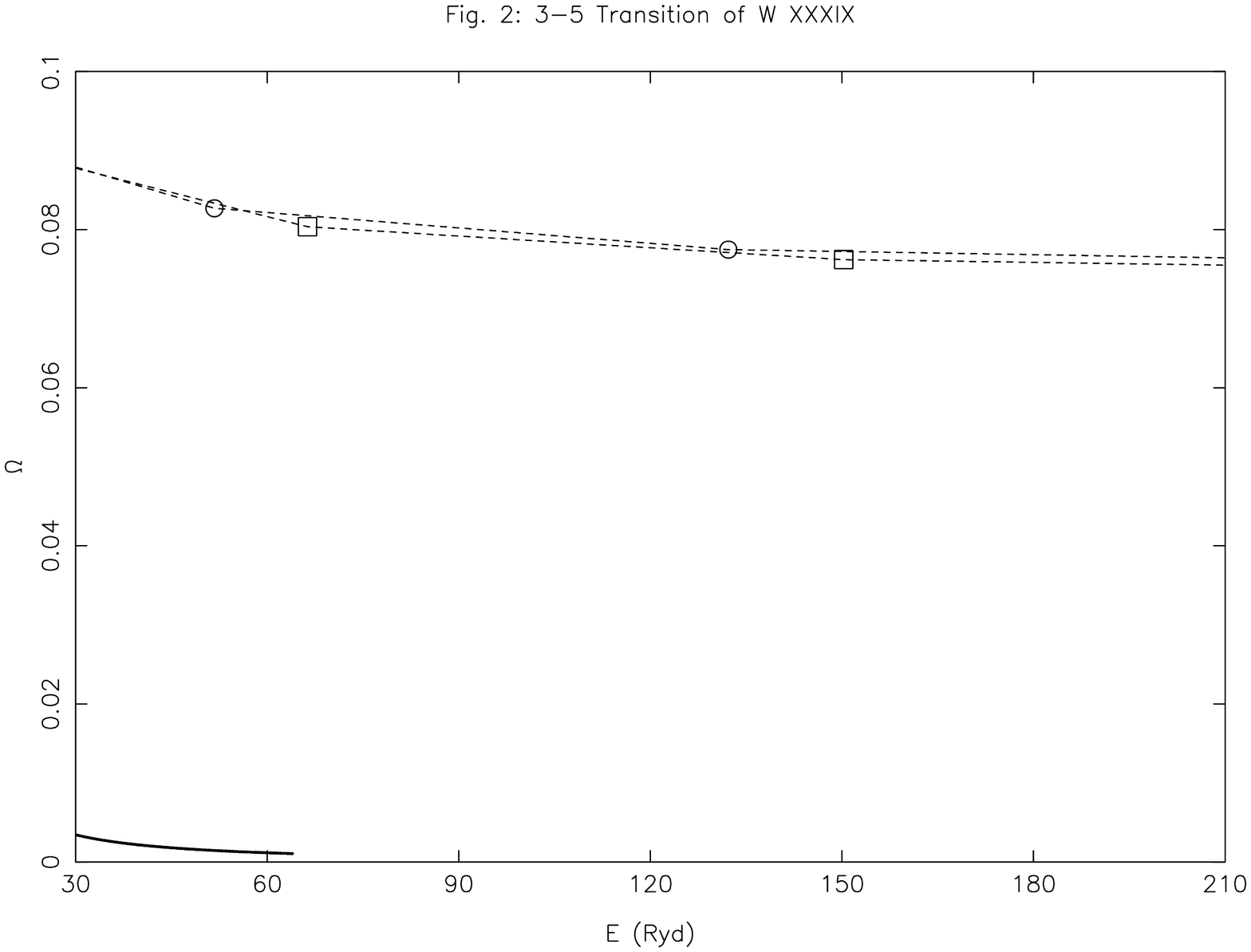}
 \vspace{-1.5cm}
\caption{Comparison of DARC and FAC values of $\Omega$ for  the 3--5 (4p$^5$4d~$^3$P$^o_1$--4p$^5$4d~$^3$D$^o_2$) transition of W~XXXIX. Continuous curve: earlier results of El-Maaref {\it et al}\, (2019) with DARC, broken curves: present results  with FAC, circles: 25 levels model and squares: 357 levels model.}
 \end{figure*}
 
 \begin{figure*}
\includegraphics[angle=-90,width=0.9\textwidth]{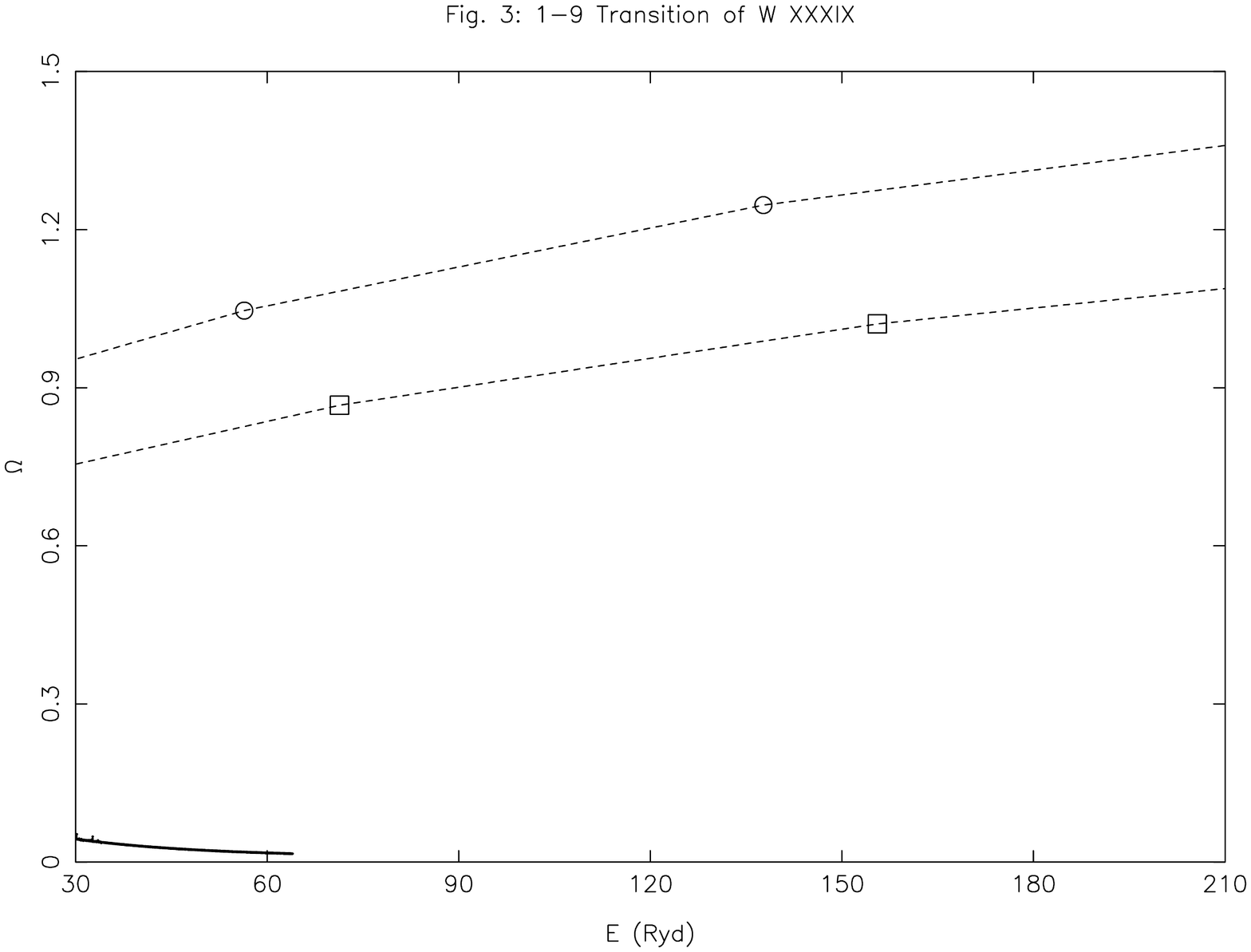}
 \vspace{-1.5cm}
\caption{Comparison of DARC and FAC values of $\Omega$ for  the 1--9 (4p$^6$~$^1$S$_0$ -- 4p$^5$4d~$^3$D$^o_1$) transition of W~XXXIX. Continuous curve: earlier results of El-Maaref {\it et al}\, (2019) with DARC, broken curves: present results  with FAC, circles: 25 levels model and squares: 357 levels model.}
 \end{figure*}
 
With FAC we have performed two sets of calculations by including the (i) 25 (FAC1) and (ii) 357 (FAC2) levels, described in section~2. In figure~1 we compare our results of $\Omega$ with those of El-Maaref {\it et al}\, (2019), obtained with DARC. It may be noted that energies shown in this and other figures are all {\em incident} although FAC lists the excited ones. For brevity, only three transitions are shown here, which are (a)  1--2 (4p$^6$~$^1$S$_0$ -- 4p$^5$4d~$^3$P$^o_0$), (b)  1--3 (4p$^6$~$^1$S$_0$ -- 4p$^5$4d~$^3$P$^o_1$) and (c) 1--5 (4p$^6$~$^1$S$_0$ -- 4p$^5$4d~$^3$D$^o_2$). This is because for 1--2 and 1--5 they have shown sudden jumps with the FAC results, both of which are forbidden transitions, and 1--3 is the only allowed one. Additionally, for clarity comparisons are shown only at energies above thresholds because our calculations do not include resonances. Our interest is in comparing $\Omega_B$, which are more important and should be comparable between the FAC and DARC calculations. 

For 1--2 and 1--5 transitions both FAC1 and FAC2\, $\Omega$ are comparable over the entire energy range and hence, do not support the findings of El-Maaref {\it et al}\, (2019), who show different $\Omega_B$ for the two models, and neither are there any jumps, as demonstrated by them. For the 1--3 transition, the two results are indeed different, but this is an {\em allowed} one, for which $\Omega$ directly depends on the f-value and the energy difference, $\Delta$E$_{i,j}$. The $\Delta$E$_{i,j}$  in FAC1 and FAC2 are comparable (11.25 and 11.75~Ryd, respectively), but the f-values are different, i.e. 0.0076 and 0.0065, respectively, and hence the differences in $\Omega$s. The corresponding f-values with GRASP are also similar (i.e. 0.0073 and 0.0064, respectively) and have no discrepancy with the result of El-Maaref {\it et al}.\,  Therefore, their results for $\Omega$ with both codes should have been comparable but are strikingly different, for all transitions,  as seen in present or their figure~1. Their results are clearly {\em underestimated}, by up to over an order of magnitude,  because they have calculated $\Omega$ with {\em limited} partial waves with angular momentum $J \le$ 9.5, which is {\em not} sufficient for the convergence. This fact has been emphasised and demonstrated in several of our papers -- see for example, Aggarwal and Keenan (2008) for Ni~XI and Aggarwal (2016) for W~LXVI. Similarly, $\Omega$ values for allowed transitions (generally) increase with energy whereas those of El-Maaref {\it et al}\, decrease, as seen in figure~1b. This is because increasingly larger number of partial waves are required with increasing energy, which they have not done. 

In figure~2 we make similar comparisons for one more transition, i.e. 3--5 (4p$^5$4d~$^3$P$^o_1$--4p$^5$4d~$^3$D$^o_2$), which is semi-forbidden and  the magnitude of $\Omega$ for this is much larger compared to those shown in figure~1. However,  the discrepancies and conclusions are the same for this  transition as are for 1--2 and 1--5, and the $\Omega$ values of  El-Maaref {\it et al}\, (2019) with DARC are clearly underestimated. Finally, in figure~3 we consider the 1--9 (4p$^6$~$^1$S$_0$ -- 4p$^5$4d~$^3$D$^o_1$) transition, which is allowed and strong with f = 1.321 in FAC1 and 1.107 in FAC2, comparable with the GRASP calculations with f = 1.3006 and 1.1062, respectively. Since $\Delta$E$_{i,j}$ is comparable in both models, the differences in $\Omega$ are proportional to the one in f-values alone. Therefore, based on the comparisons shown in figures~1--3 we conclude that the reported $\Omega$ results by El-Maaref {\it et al}\, are highly underestimated (by up to over an order of magnitude), and for both forbidden as well as allowed transitions, because of the inclusion of limited range of partial waves, and hence the non-convergence. Their results with FAC are also incorrect because of the sudden jumps in $\Omega$ behaviour, which we neither observe nor expect. Apart from this, there are other deficiencies in their work which we discuss below.

As is well known and has also been emphasised by El-Maaref {\it et al}\, (2019), tungsten is one of the most useful material for fusion reactors, and atomic data for its ions are required for a variety of studies. Since the temperatures in fusion plasmas are very high (in the range of $\sim$ 10$^6$ to 10$^8$~K or equivalently 6.3 to 633~Ryd), calculations for $\Omega$ need to be performed up to very high energies, and the one done by them for up to 64~Ryd is of no use. Since the highest threshold considered in their work (level 357) is $\sim$80~Ryd, the calculations for $\Omega$ should be performed for up to, at least, 700~Ryd. Unfortunately, their $\Omega$ values can also not be reliably extrapolated to higher energies, and if attempted then it may lead to faulty results, as demonstrated and concluded in some of our work on Be-like ions, see for example, Aggarwal and Keenan (2015a,b) and Aggarwal {\it et al}\, (2016). Similarly, the resonances demonstrated by them for a few transitions are only useful for calculating the {\em effective} collision strengths ($\Upsilon$), obtained after integration over an electron velocity distribution function, mostly {\em Maxwellian}. Since their energy resolution is rather coarse (0.02~Ryd or equivalently  3160~K), this will not yield accurate and reliable results, as has recently been discussed and demonstrated by us (Aggarwal 2018a,b) for two F-like ions. Finally, they have reported their results for only 57 transitions, out of the possible 46~971 among the 307 levels considered, i.e. $\sim$0.1\%. For modelling of plasmas results for a complete model are desired and therefore their reported data are not of much practical use.

\section{Conclusions}

In this short communication, we have demonstrated that the collisional data recently reported by El-Maaref {\it et al}\, (2019) for transitions in Kr-like W~XXXIX are highly underestimated for all transitions, irrespective of their types, such as forbidden or allowed. This is because they have considered a very limited range of partial waves with $J \le$ 9.5, highly insufficient for the convergence of $\Omega$. Similarly, their reported results are for about 0.1\% of the possible transitions among the considered 307 levels, and are therefore of limited practical use, even if correct. Therefore, a more reliable, accurate and complete set of collisional data for this ion is highly desirable. Similarly, there is scope for improvement in the accuracy of the results reported for energy levels and radiative rates.

It has been suggested by several authors in the past  (see for example, Aggarwal and Keenan 2013,  Chung {\it et al}\, 2016 and Aggarwal 2017) that  to assess the reliability and accuracy of atomic data  independent calculations should be performed, either with the same or a different method/code, which El-Maaref {\it et al}\, (2019) have done. However, such calculations are only useful if performed diligently and carefully, and differences, if any, are resolved rather than ignored. Similarly, merely comparing data for a few levels or transitions may result in faulty conclusions, as has been done by El-Maaref {\it et al},\,  and earlier by Tayal and Sossah (2015), as explained in our paper on Mg~V (Aggarwal  and Keenan 2017). Therefore, we will like to emphasise again that the reliability of any calculation does not (much) depend on the (in)accuracy of the method/code adopted, but on its implementation. An incorrect application of a code may lead to large  discrepancies and faulty conclusions. Finally, for the benefit of the readers we will like to note that some other works reported by El-Maaref  and co-workers on other ions are deficient and erroneous for similar reasons, as recently demonstrated and explained by us for W~XLV (Aggarwal 2019a), Mn~X (Aggarwal 2019b), and Sc~VI (Aggarwal 2019c,d). 


\end{document}